\documentclass{jpp}
\usepackage{graphicx}
\usepackage{epstopdf, epsfig}
\usepackage{hyperref}

\shorttitle{Interaction of parallel Alfv\'en waves }
\shortauthor{K. Daiffallah}

\title{Electron acceleration from the interaction of three crossed parallel Alfv\'en waves}

\author{K. Daiffallah \aff{1}
 \corresp{\email{k.daiffallah@craag.dz}}}

\affiliation{\aff{1} Centre de Recherche en Astronomie, Astrophysique et
Geophysique CRAAG (Observatory of Algiers), Division Astrophysique Solaire, Route de l'Observatoire, BP 63, Bouzar\'eah 16340, Algiers, Algeria}

\begin{document}
\maketitle
\begin{abstract}
We study the non-linear interaction of three parallel Alfv\'en wave packets (AWP) in an initially uniform plasma using 2.5D particle-in-a-cell (PIC) numerical simulations. We aim to help to explain the observation of suprathermal electrons by multiple Alfv\'en waves collision in regions where these waves are trapped like IAR (Ionospheric Alfv\'en Resonator), Earth radiation belts or coronal magnetic loops. 
In the context of APAWI process that have been described by \citet{mottez12, mottez15}, the interaction of two parallel Alfv\'en waves (AW) generates longitudinal density modulations and parallel electric field at the APAWI crossing region that can accelerate particles effectively in the direction of the background magnetic field.
Our simulations show that when a third parallel AWP of different initial position arrives at the APAWI crossing region,  it gives rise to a strong parallel electron beams ($V \sim 5-7 V_{Te}$) at longitudinal cavity density gradients. We suggest that velocity drift from outgoing AW generates interface waves in the transverse direction, which allows propagating waves to develop parallel electric fields by phase mixing process when $k_{\perp}^{-1}$ of the wavy density gradient (oblique gradient) is in the range of the electron inertial length $c/\omega_{p0}$.

\end{abstract}

 
\section{Introduction}\label{intro}
Nonlinear interactions of Alfv\'en waves play a substantial role in the heating and acceleration of  the solar corona and in the auroral acceleration of particles in Earth magnetosphere. The collision between counter-propagating Alfv\'en waves has been investigated in term of MHD and kinetic plasma by a number of authors \citep[e.g.,][]{zhao11,howes113, nielson213, pezzi17, howesjul18, verniero18}. However, these studies have been focused mainly on the dynamics of turbulent energy transfer to the small perpendicular scales range.

Acceleration by Parallel Alfv\'en Waves Interactions (APAWI) described in  \citet{mottez12,mottez15} is a process of nonlinear interaction of two Alfv\'en wave packets propagating mainly in opposite directions, in an initially uniform plasma, with their wave vectors parallel to the ambient magnetic field. The APAWI process causes a significant modulation of the plasma density giving rise to cavities and electron depletion at the waves crossing region. Alfv\'en waves with the same circular polarization can generate a parallel magnetic gradient and electric field accelerating electrons in the direction of ambient magnetic field. 
APAWI can take place in regions associated to trapped Alfv\'en waves like IAR (Ionospheric Alfv\'en Resonator) and radiation belts in Earth, or in magnetic loops in solar corona (see \citet{mottez15} for discussions). However, in such plasma configuration, waves can suffer from several parallel reflections giving rise to more complex interactions. Therefore, unlike previous studies \citep{mottez12,mottez15} were only two sinusoidal waves or two wave packets were considered, we consider the interaction of a larger number of waves. For simplicity and to understand the basic physics of particle acceleration by multiple Alfv\'en wave collisions,  we have carried out simulations of interaction of three parallel Alfv\'en waves. 

We use a particle in cells (PIC) numerical simulation code \citep{mottez98} to study the interaction of three parallel Alfv\'en waves packets in an initially homogeneous plasma. First, two wave packets interfere, and behave according to the APAWI process.  Then, a third Alfv\'en wave packet arrives at the crossing region (inhomogeneous plasma) of the two initial Alfv\'en waves, which, as we will show, is  equivalent to the interaction of a single Alfv\'en wave packet with an inhomogeneous plasma.

A phase-mixing process occurs when an Alfv\'en wave propagates through a density inhomogeneity that is transverse to the background magnetic field \citep{priest83}.
Initially, the Alfv\'en wave propagates in a parallel direction ($k_{\perp}=0$) with a plane surface front. However, when it hits a perpendicular gradient of density like cavity (density depletion), the wavefront is distorted (phase mixing) because of the fast velocity propagation in the low density region than in the outside according to the Alfven velocity $V_A=B_0/(\mu_0\rho_m)^{1/2}$, where $\rho_m$ is the mass density and $B_0$ is the background magnetic field. Therefore the wavefront becomes oblique to the parallel magnetic field and develops a perpendicular wave vector $k_{\perp}$ along the transverse density gradients as well as small perpendicular scales. When this transverse scale $k_{\perp}^{-1}$ is of order of the electron inertial length ($c/\omega_{p0}$) in the inertial regime where $\beta << m_e/m_i$ ($\beta=2 \mu_0 p/B_0^2$ is the ratio kinetic pressure over the magnetic pressure), the AW develops a parallel electric field in strong density gradient regions which accelerates electrons in the parallel direction \citep{genot99}. The parallel electric field $E_{||}$ is given by the general equation \citep{goertz85}:

\begin{equation}
\frac{E_{||}}{E_{\perp}} = -\frac{c^2k_{\perp}k_{||}/\omega_{p0}^2}{1+c^2k_{\perp}^2/\omega_{p0}^2},
\end{equation}

where $\omega_{p0}$ is the electron plasma frequency, $c$ is the speed of light.

For plasma cavity of density $n$, the parallel electric field is obtained by replacing $k_{\perp}$  with $(1/n) \partial_{\perp} n$, where $\partial_{\perp}$  is the spatial derivative of the density in the transverse direction to the parallel magnetic field \citep{genot2000}.

In the kinetic regime  where $ m_e/m_i<< \beta << 1$, a parallel electric field is generated when the $k_{\perp}^{-1}$ reaches the ion gyroradius length ($\rho_i=m_i v_{\perp}/e B_0$) \citep[see][]{bian11}.  

Some electrons are accelerated to velocities larger than the wave phase velocity. Therefore, they escape from acceleration regions and create a beam. In some case this beam becomes unstable when it interacts with the plasma giving rise to a beam-plasma instability which evolves to small-scale electrostatic structures where particles may be trapped. Electron or phase space holes are an another kind of instability structures that can also stop the acceleration process by trapping particles in their electric field \citep{mottez01}. They are associated to a positive potential perturbation and corresponds to low density of particles trapped by initially low amplitude wave. We identify these structures in the phase space in the form of vortex of dimensions $\delta X$ and $\delta V_X$, where the middle of a vortice is centered at phase velocity and propagating with this velocity.

The outline of the paper is as follows: The simulation setup and parameters are described in section 2. Numerical simulation of crossing of three parallel Alfv\'en wave packets is presented in section 3. Two cases are discussed in this section: (i) Alfv\'en waves with the same polarization, (ii) Alfv\'en waves with different polarizations and initial positions. Finally, discussion and conclusion are presented in the last section.

 
\section{Simulation setup and parameters}\label{intro}

All physical variables in the code are dimensionless e.g., the frequency $\omega$ is normalized to electron plasma frequency $\omega_{p0}$, charge densities $n$ to the electron density $n_0$, the masses to the mass of the electron $m_e$, the charges to the electron charge $e$, the velocities to the speed of light $c$. The dimensionless distances and wave vectors are given by the ratios  $c /\omega_{p0}$ and $\omega_{p0}/c$ respectively. The dimensionless background magnetic field $B_0$ is given by the ratio $\omega_{ce}/\omega_{p0}$, where $\omega_{ce}$ is the electron cyclotron frequency.  The normalized electric field is $ E \omega_{ce} /\omega_{p0} cB$ 

The initial conditions consist of a uniform plasma with identical electron and ion temperatures $T_e=T_i$.

The initialization of the AW and their polarization is based on the resolution of the dispersion equation in the context of bi-fluid theory of the cold plasma.  In the case of parallel propagation to the background magnetic field, the fourth order polynomial dispersion equation has four roots; the two with highest frequency are rejected, where the two with lowest frequency corresponds to the right-hand (RH) and left-hand (LH) circularly polarized AW. 
Then, the other perturbations are set depending on the choice of polarization solution (RH or LH). More details are given in the appendix in  \citet{mottez2008}. It is important to note that the phase velocity of the RH wave is larger than the Alfven velocity, whereas it is smaller in the case of LH wave.

The magnetic field can be written as the sum of the uniform $B_0$ along the $X$ axis, and the sum of the waves magnetic fields labeled $i$, with $i=1,2,3$.
\begin{eqnarray} 
B_X &=& B_0 \label{eq_B}\\
\nonumber
B_Y &=& \sum_i B_{iY} \cos(\omega t - kX +\phi_{B_{iY}})\\   
\nonumber
B_Z &=& \sum_i B_{iZ} \cos(\omega t - kX +\phi_{B_{iZ}}). 
\end{eqnarray}
We write the phase relations in the form 
\begin{equation}
\phi_{B_{iZ}} = \phi_{B_{iY}} + {n_i \pi}/{2}
\end{equation}
with $n_i =\pm 2$ for linear polarization, $n_i=+1$ for right-handed waves, 
$n_i=-1$ for left-handed waves. With circularly polarized waves, $B_{iY}=B_{iZ}$ and we write simply $B_i$.

 The simulation box size is $L_X \times L_Y= N_X \Delta X \times N_Y \Delta Y$. The cell size is $\Delta X=\Delta Y = V_{Te}= \lambda_{De}$, where  $\lambda_{De}$ and $V_{Te}$ are the Debye length and the electron thermal velocity respectively. The simulations are done in two dimensions ($N_X=4096$, $N_Y=64$).
The total duration of the simulations is $16384 \times \Delta t = 1638.4$, where $\Delta t$ is the time step defined by $\Delta t =0.1$. For all simulations, the ion to electron mass ratio is reduced to $m_i/m_e= 100$.

Following \citet{mottez15}, the background magnetic field and the electron
thermal velocity are $B_0=0.8$ and $V_{Te}=0.1$ respectively. With these plasma characteristics, we find $(m_i/m_e)\beta=(V_{Te}/B_0)^2=0.016$  which corresponds to a plasma in the inertial regime where $(m_i/m_e) \beta << 1$. This are the case of large regions in the high altitude auroral zone and inner
solar corona. The size of the simulation box is  $L_X=N_X \times \Delta X=409.6$ and  $L_Y=N_Y \times \Delta Y= 6.4$.

We consider that one wave packet (AWP) is a sum of 16 sinusoidal waves where the maximum of magnetic amplitude is located at it initial position $X_0/L_X$. The waves have a right handed circular polarization (RH) or a left handed circular polarization (LH). The wavelengths are $\lambda=\lambda_0/m$, where $\lambda_0=L_X=409.6$ and the wave number $m$ varies from 1 to 16. The Alfv\'en velocity is $V_{A0}=0.08$, where the phase velocities varies from  0.2029  for the shortest wave to 0.0855  for the longest one in the case of (RH) AWP, and varies from 0.02846 to 0.07362 in the case of (LH) AWP. 

The ratio $\omega/\omega_{ci} \approx 0.1639$ for (RH) waves of large wavelengths ($m=1$). This corresponds to the high-frequency part of the MHD Alfv\'en waves ($\omega/\omega_{ci} \ll 1$). With waves of short wavelengths ($m=16$), the ratio $\omega/\omega_{ci} \approx 6.225$ do not corresponds with purely MHD Alfv\'en wave, but with an electron cyclotron wave (or Whistler wave) which is situated on the upper frequency part of the same dispersion relation branch. All the (LH) waves frequencies are lower than the ion cyclotron frequency $\omega_{ci}$ ($\omega/\omega_{ci} \approx 0.1412$ for $m=1$ and $\omega/\omega_{ci} \approx 0.8731$ for $m=16$). We are still in the high frequency range of the MHD Alfv\'en waves branches.

\begin{figure} 
\center
\vspace{0.05\textwidth}
\includegraphics[width=6.5cm]{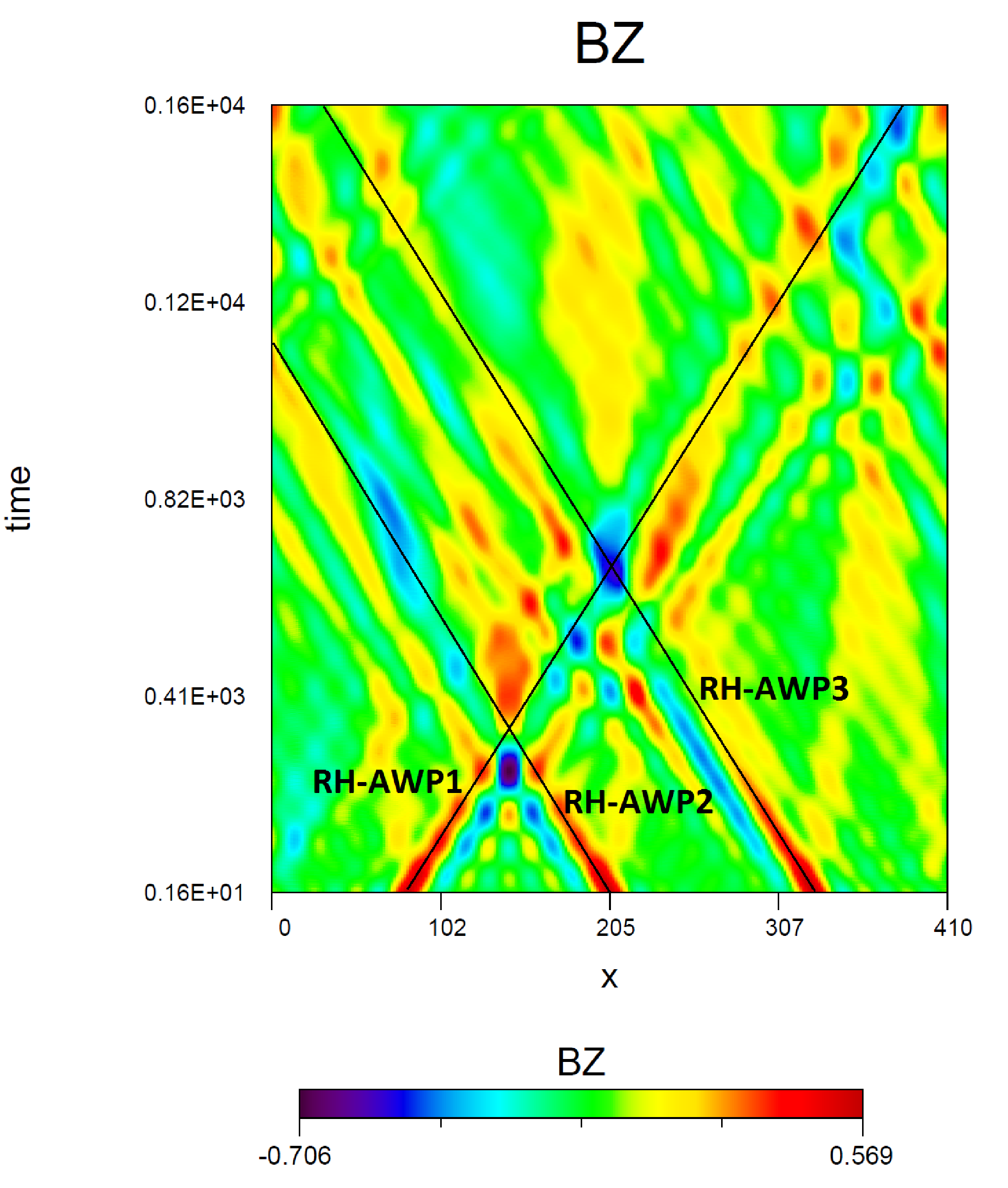}
\hspace{0.03\textwidth}
\includegraphics[width=6.3cm]{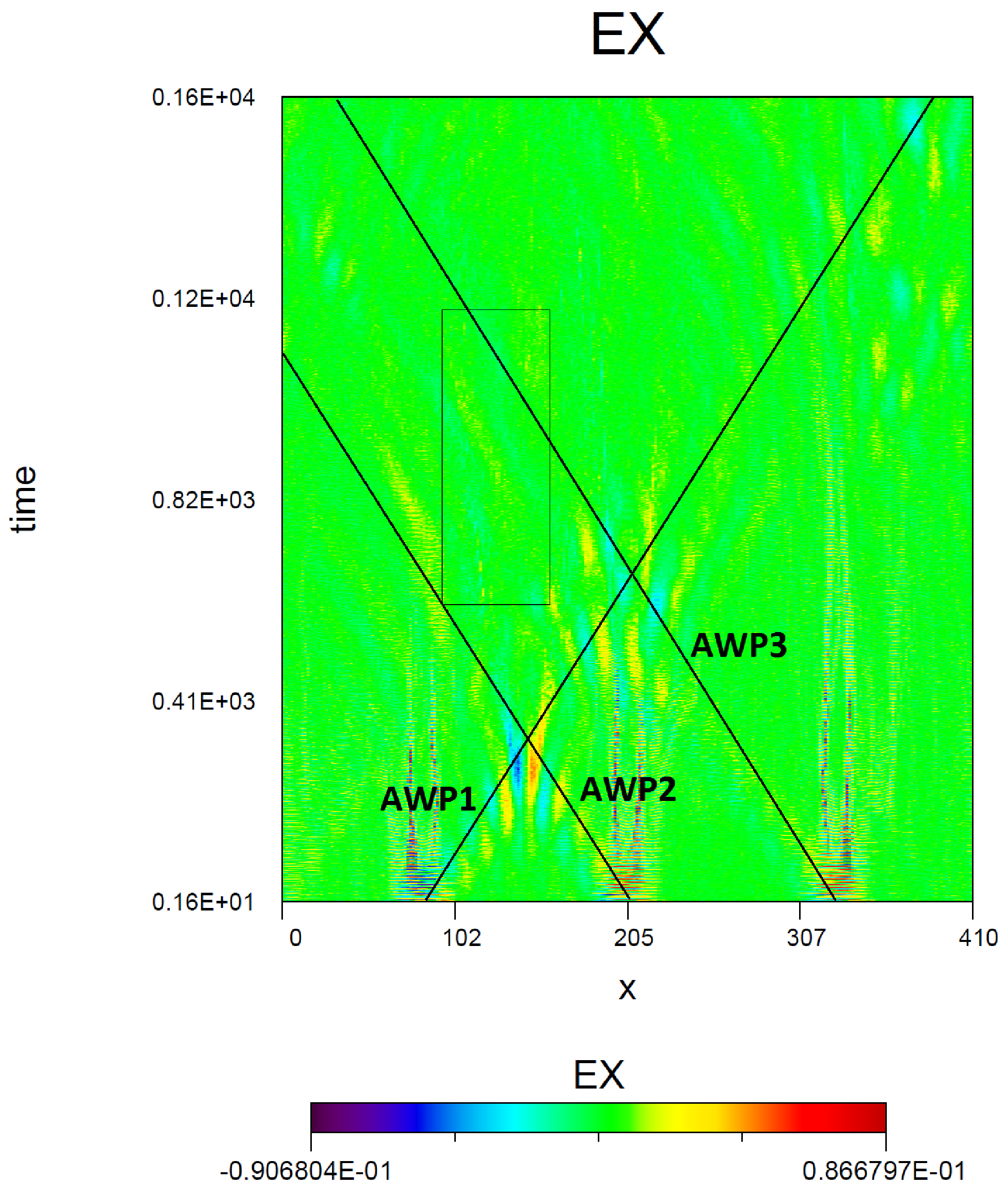}
  \caption{Run (a): The temporal variation of the components $B_z(X,t)$ (left panel) and $E_X(X,t)$ (right panel) as a function of the
    parallel coordinate $X$. The oblique lines in both panels show the time-distance path of the three wave packet centres (RH-AWP). The rectangle in the right panel circumscribes the electric fields that can emerge from non-APAWI process. }
\label{fig2cbzc}
\end{figure}

According to \citet{buti00}, the left handed (LH) AWP are unstable and they can collapse or change the polarization in $\beta \ll 1$ regime, while the (RH) waves are more stable in this regime.
Since we are propagating (LH) Alfv\'en waves in our simulations we have checked this possibility. A simulation of the propagation of a single (LH) wave with initial amplitude of $\delta B/B_0=0.1$ shows a localized wave packet as in the case of (RH) wave. Both AWP propagate without changing their direction and they are not destroyed by the dispersion.
  
We can notice that the initial Alfv\'en waves in \citet{buti00} have a large value $\delta B/B_0=0.5$, where the collapsed Alfv\'en wave packets seem to stabilize around a amplitude of $\delta B/B_0=0.2$. In our study, the amplitude is low, and this explains why we do not observed the instability described in  \citet{buti00}.


\section{Simulation of crossing of three parallel Alfv\'en wave packets}\label{sec:types_paper}

We present a series of numerical simulations Runs (a),(b),(c) and (d). The table \ref{tab1} shows the runs and initial conditions.
In these simulations, the number of particles per cell is fixed to $N=50$ or $N=100$ for simulations. The first wave packet propagates toward increasing values of $X$ (downward direction D), whereas the second one propagates toward decreasing values of $X$ (upward direction U). The third wave packet propagates upward.


\subsection {Alfv\'en waves with the same (RH) polarization}\label{subsec1}
\subsubsection{ Simulation (a)}

Simulations (a) shows the interaction of two Alfv\'en wave packets followed by the passage of a third another wave packet. The initial positions of the three wave packets are $X_0/L_X=0.2, 0.5$ and  0.8 respectively. Their amplitudes and polarization (RH) are the same. 
The simulation (a) is shown in Figure \ref{fig2cbzc}. The left panel shows the
interaction of three wave packets of the same polarization (RH) through the
temporal evolution of the component $B_Z(X,t)$. The oblique lines represent the time-distance path of the three wave packet centres (RH-AWP). In this simulation, the crossing of the two first wave packets occurs from $t\approx 205$ at the position $X\approx 145$. The third packet interact with the crossing region from $t \approx 410$ at the position $ X \approx 205$. In fact, the third wave will interact initially  with one of the initial (RH) waves before its passage through the crossing region. We can observe that the two initial and the third wave packets continue to propagate in their original direction. They are not destroyed after their interactions.

\begin{table}
\begin{center}
\def~{\hphantom{0}}
\begin{tabular}{l c c c c }        
Run & polar  & $\delta B/B_0$ & init positions $X_0/L_X $ & Observations  \\[6pt]   
   (a)& RH-RH RH & 0.05-0.05 0.05  & 0.2-0.5 0.8 & $f_e$ vortices, electron and ion beams \\
   (b)& RH-RH RH & 0.1-0.1 0.1  & 0.2-0.5 0.8 & strong $f_e$ vortices, electron and ion beams \\
   (c)& RH-LH LH  & 0.05-0.05 0.05 &  0.2-0.5 0.8 & $f_e$ vortices, ion beams   \\     
   (d)& RH RH-LH & 0.05 0.05-0.05 &  0.2 0.5-0.8 & $f_e$ vortices, ion beams   \\
     \end{tabular}
\caption{Simulation runs and initial conditions. The dash between two
  polarizations in the second column indicates the interaction of the initial AWP. The observations concern the three AWP crossing region. }            
\label{tab1} 
\end{center}
\end{table}

We have to notice that  because of the periodicity of the box in $X$
direction, one of the initial waves $(X_0/L_X=0.5)$ which has already
interacted with the first one $(X_0/L_X=0.2)$ reappears in the right side of
the box and interacts once again with the first one at  $X\approx 358$ leading
to what we call "artificial" interactions mainly visible at the end of
simulations. 

The right panel of Figure \ref{fig2cbzc} shows the temporal evolution of the parallel electric field $E_X(X,t)$ for simulation (a). A first quasi-stationary parallel electric field is observed at $X\approx 145$ and it is associated to the collision between the two initial Alfv\'en waves according to APAWI process. While the APAWI electric field vanishes at  $t\approx 546$,  a weaker ones emerge mainly for $102 \le X \le 145$ from  $t\approx 600$  and they seem follow the propagation of the third AW. These non-APAWI electric fields are faintly visible inside AWP1-AWP2 interaction region circumscribes by the rectangle.

A second APAWI quasi-stationary parallel electric field is observed at $ X \approx 205$ from $t\approx 410$ which is associated to the collision between the third AWP with the first initial one $(X_0/L_X=0.2)$.  

\begin{figure} 
\center    
\vspace{0.01\textwidth}
\includegraphics[width=6.8cm]{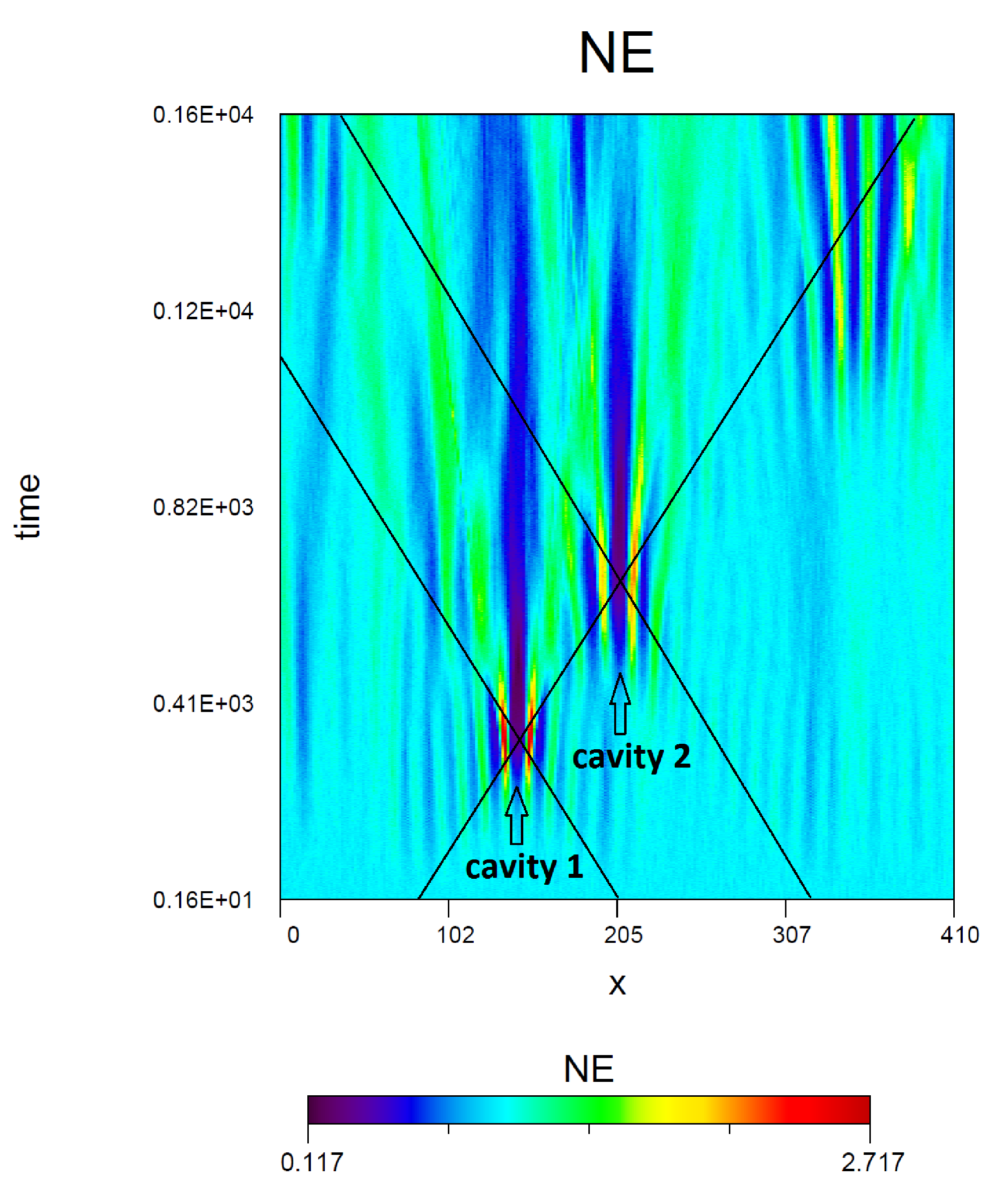}  
\hspace{0.03\textwidth}
\includegraphics[width=6cm]{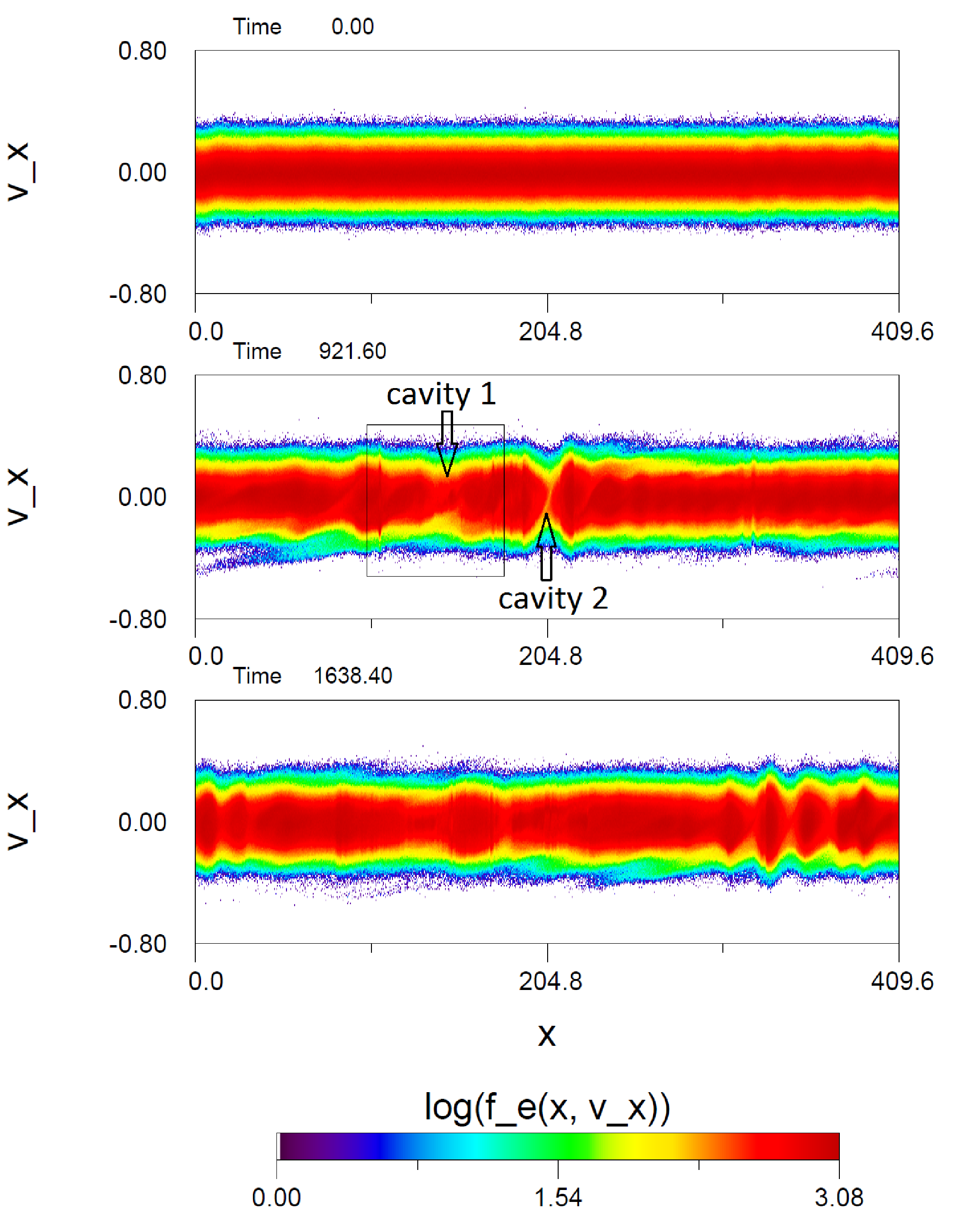}   
  \caption{Run (a): The left panel shows the temporal variation of the electron density $N_e(X,t)$, where cavity 1 and cavity 2 are the location of the two emerged density depletions from APAWI process. The right panel shows snapshots from $t = 0$ to $t = 1638.4$ of the electron parallel 
distribution function $f_e(X, V_X)$ in logarithmic scale. The rectangle in the snapshot $t = 921.6$ circumscribes the AWP3-cavity 1 interaction region.}
\label{12ver6}
\end{figure}

Simulations from \citet{mottez12, mottez15} have shown the emergence of plasma density perturbation and cavities when two Alfv\'en waves interact (APAWI). 
The left panel of Figure \ref{12ver6} shows the temporal evolution of the electron density $N_e(X,t)$ for simulation (a).   
As observed by \citet{mottez12, mottez15}, a large part of density modulations that emerge from waves crossing does not dissipate and persist during all our simulations. The crossing of two initial right-hand Alfv\'en wave packets shows a high amplitude depletion of the electron density at the locus of the interaction between the two waves at $X \approx 145$ (cavity 1). A second cavity observed at $X \approx 205$ results from the interaction of the third wave packet with the first one (cavity 2).

The right panel in Figure \ref{12ver6} shows the parallel distribution function of electrons $f_e$. At $t=921.6$, we observe electron phase space holes in form of vortices ($f_e$ vortices) inside the rectangle where the AWP3 crosses the cavity 1 (from $X\approx 102$ to $X\approx 205$). The typical size of these structures is about $\delta X \le 15$. An electron beam of velocity $\sim 5 V_{Te}$ emerges inside the rectangle in the distribution function near the position $ 110 \le X \le 128$. 

\begin{figure} 
\center
\vspace{0.05\textwidth}
\includegraphics[width=6cm]{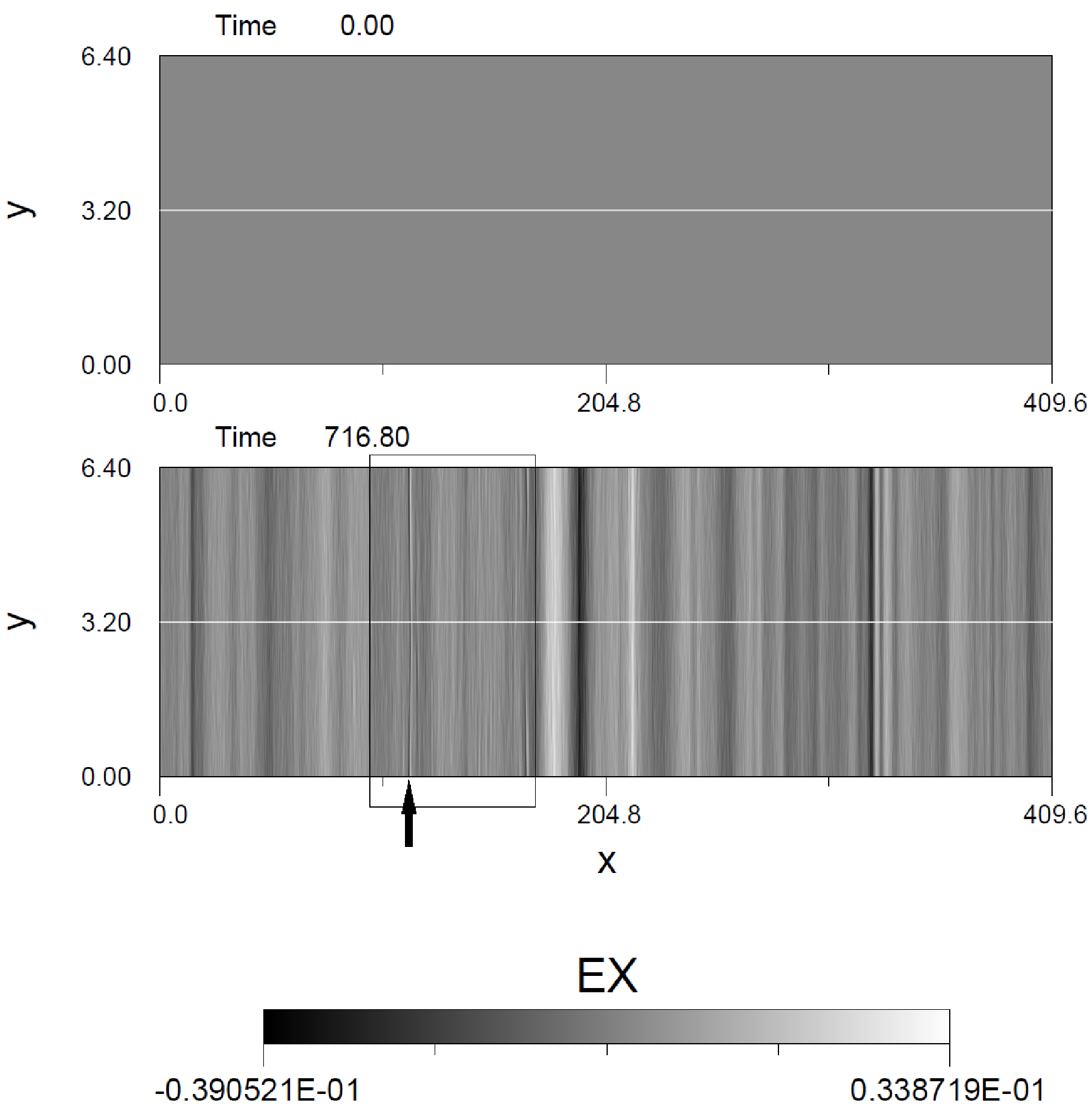} 
\hspace{0.03\textwidth}
\includegraphics[width=6cm]{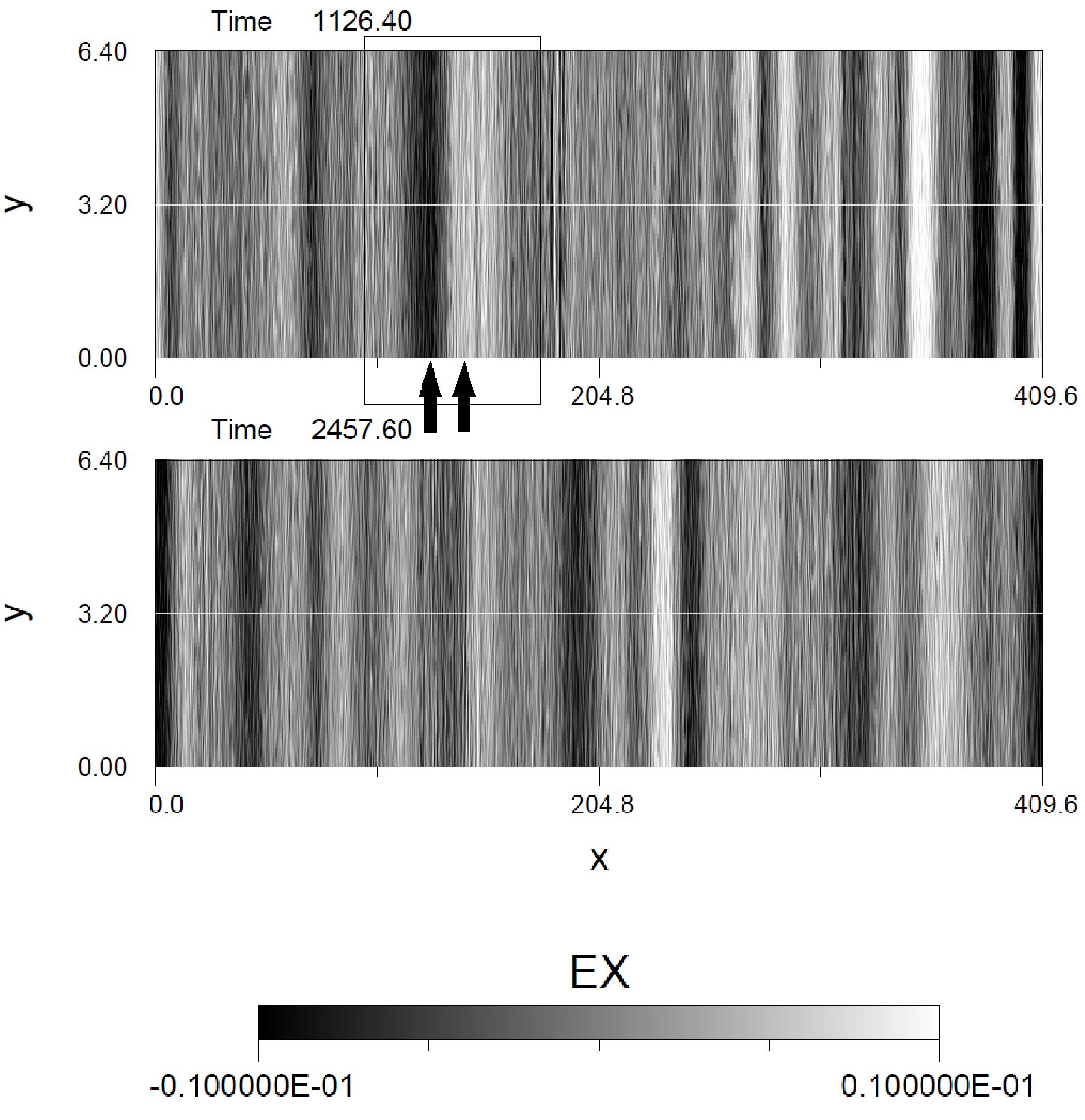} 
  \caption{Run (a): Snapshots from $t = 0$ to $t = 2457.6$ of the parallel electric field  $EX(X,Y)$. The black rectangle in snapshots $t=716.8$ and $t=1126.4$  circumscribes the region where AWP3 crosses the cavity 1. The black arrows show the location of the strong parallel electric fields that arise inside this region.}
\label{fig2.5}
\end{figure}

In the left panel of Figure \ref{fig2.5}, a very narrow size electric field ($|E| \sim 0.023$) is observed in the snapshot $t=716.8$ at the position $X \approx 115$ (see the black arrow). In the right panel of Figure \ref{fig2.5}, a more large fibril electric fields of slightly less strong strength ($|E| \sim 0.011$) are observed in the snapshot $t = 1126.4$ at $X\approx 125 $ and $X \approx 143 $ respectively (see black arrows). The small scale electric field at $t=716.8$ is visible at the bottom of the rectangle in right panel of Figure \ref{fig2cbzc}, where the large size ones are faintly visible from the middle to the top of the rectangle. It is interesting to observe that the electron beam emerges between the locations of the small size electric field and the large ones.  

\begin{figure} 
\center
\vspace{0.05\textwidth}
\includegraphics[width=5.9cm]{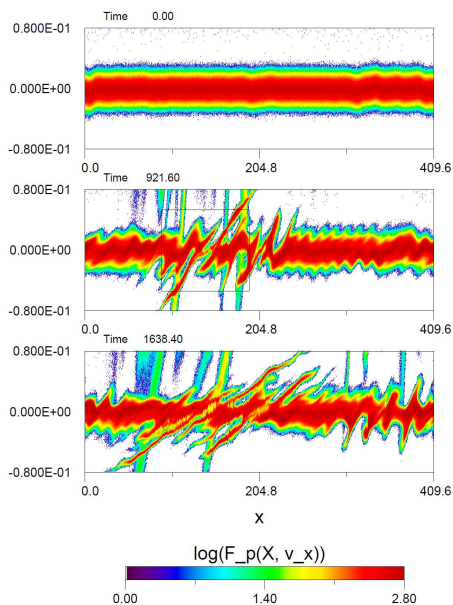}  
  \caption{Run (a): Snapshots of ion parallel distribution function $f_p(X, V_X)$ in logarithmic scale.}
\label{fig2.8}
\end{figure}

Ion beams are also observed in the region where the third AWP crosses the cavity 1 at $t=921.6$ (Figure \ref{fig2.8}).

\subsubsection{ Simulation (b)}

\begin{figure} 
\center
\vspace{0.05\textwidth}
\includegraphics[width=6.8cm]{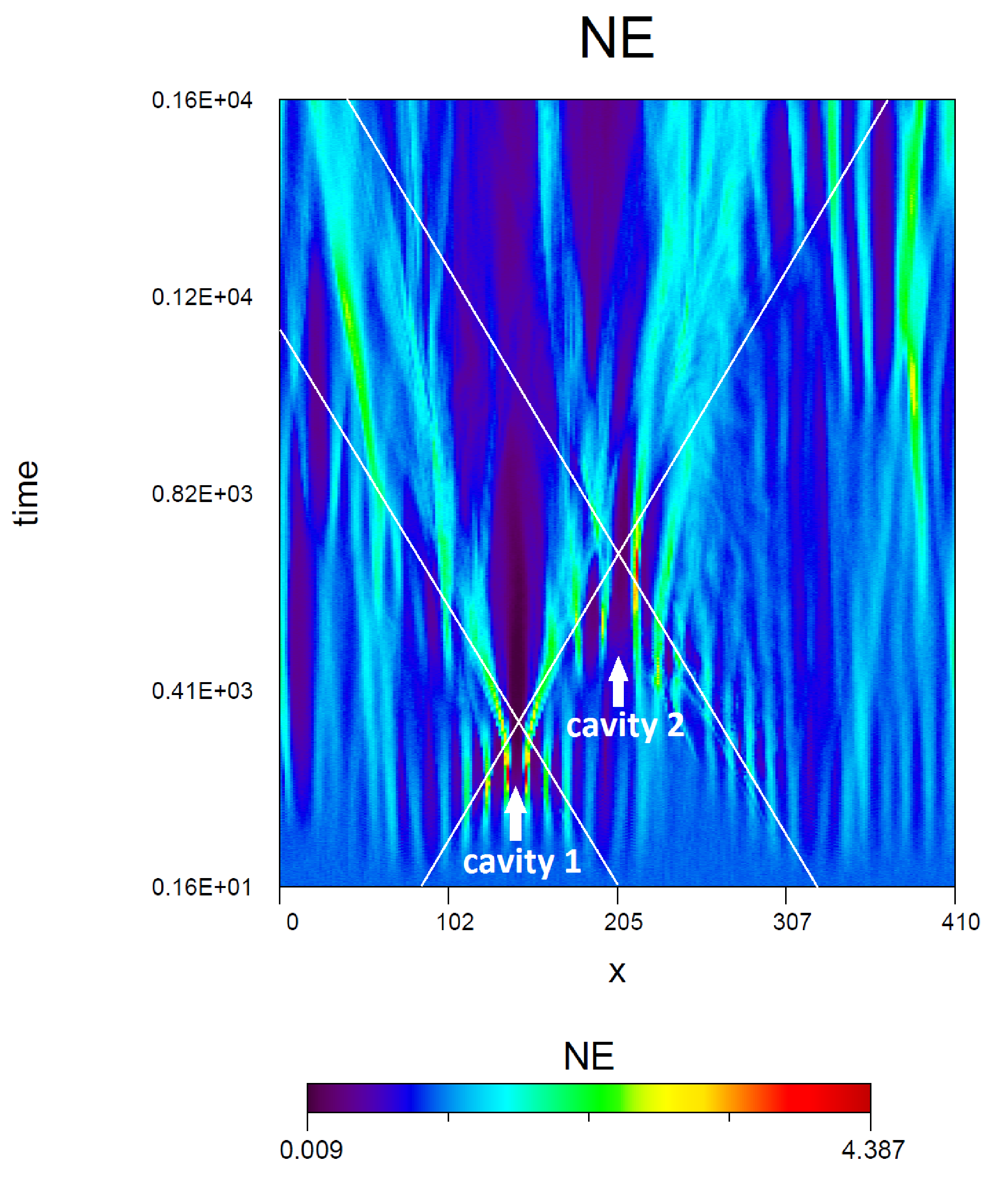}
\hspace{0.03\textwidth}
\includegraphics[width=6cm]{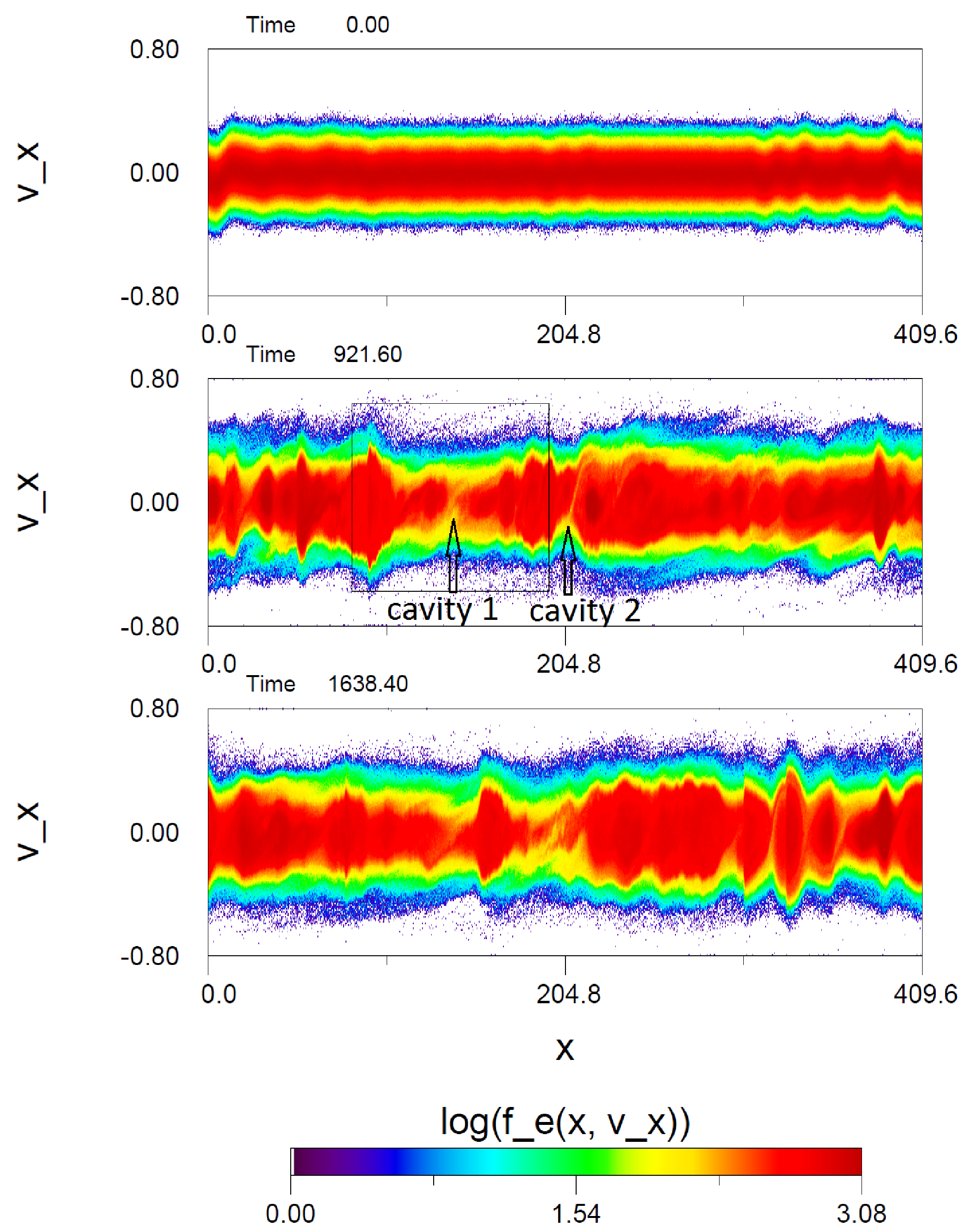}
  \caption{Run (b): The left panel shows the temporal variation of the electron density $N_e(X,t)$.The oblique white lines indicate the time-distance paths of the wave packet centres. The right panel shows snapshots from $t = 0$ to $t = 1638.4$ of the electron parallel distribution function $f_e(X, V_X)$ in logarithmic scale.}
\label{fig33b}
\end{figure}

The simulation (b) is the same as the simulation (a) except that the waves amplitude is increased two times ($\delta B/B_0=0.1$).
In the left panel of Figure \ref{fig33b} we observe a larger cavity depletion amplitude at both positions $ X \approx 145$ and  $ X \approx 205$. This is the consequence of large amplitude propagating waves in comparison to the simulation (a). 
Larger amplitude electron vortices associated to the AWP3-cavity 1 interaction region  were observed in the parallel distribution function of electrons in right panel of Figure \ref{fig33b}. This may be the direct consequence of the AW higher amplitude and instabilities which catch more particles of the core distribution and the beams \citep{mottez01}. A tiny and scattered electron beams (Velocity $\sim 7 V_{Te}$) emerge from AWP3-cavity 1 crossing region ($93 \le X  \le 205$).

The left panel of Figure \ref{figexb} shows the temporal variation of parallel electric field $EX(X,t)$.  We can observe a non propagative high frequency oscillations at the initial positions of the wave packets. These noises background are caused by an imperfect initialization of the wave modes. They are visible also in the right panel of Figure \ref{fig2cbzc}.  
In comparison to the simulation (a), a larger and stronger APAWI quasi-stationary electric fields are observed at crossing regions of two Alfv\'en waves. Another electric fields are observed at the region where the third Alfv\'en wave crosses the first cavity depletion from $t\approx 600$ (the black rectangle region). The stronger ones are clearly visible at the top of the rectangle for $1000\le t \le 1200$. However, the last parallel electric field is more extended in space and it is not stationary as the APAWI ones. 

The parallel electric fields associated to AWP3-cavity 1 crossing region are visible in the right panel of Figure \ref{figexb} at the snapshot $t=1126.4$ (see the black arrows). They are more stronger than those of the simulation (a) ($|E| \sim 0.038$) and they show also a fibril structure.

\subsection {Alfv\'en waves with different polarizations and initial positions}\label{subsec2}
\subsubsection {Simulations (c) and (d)}

The simulation (c) concerns the interaction of a left-hand waves (LH) with two initial Alfv\'en waves of different polarization (RH-LH). The initial positions of the waves are the same as in simulation (a).  The simulation (d) is for the interaction of a right-hand waves (RH) with two initial waves of different polarization (RH-LH). Unlike simulations (a),(b) and (c), the two initial wave packets in simulation (d) are initialized at  $ X_0/L_X=0.5$ and $X_0/L_X=0.8$, respectively, whereas the third wave packet started at the position $ X_0/L_X=0.2$.  In both simulations (c) and (d), we observe a weaker cavity depletion amplitude and a weaker parallel electric fields strength in comparison to the simulation (a). 


\subsection {Phase mixing and longitudinal density gradient }\label{subsec2}

 Given that, an interesting question emerges: can phase-mixing process explains the observation of the non-APAWI parallel electric fields from the passage of the third Alfv\'en wave packet through cavity 1 region (density inhomogeneity gradients)? 

Phase-mixing has been associated to density depletion or cavity (plasma under-density) in the auroral regions \citep{genot99,genot2000, mottez11} or more recently in interplume regions \citep{daiffallah17}. This mechanism has been associated also to density bump or plasma over-density in the solar coronal loops \citep{tsi07,tsi11, tsi12, tsi16} or in a solar coronal hole \citep{wu03}. 

\begin{figure} 
\center
\vspace{0.05\textwidth}
\includegraphics[width=7cm]{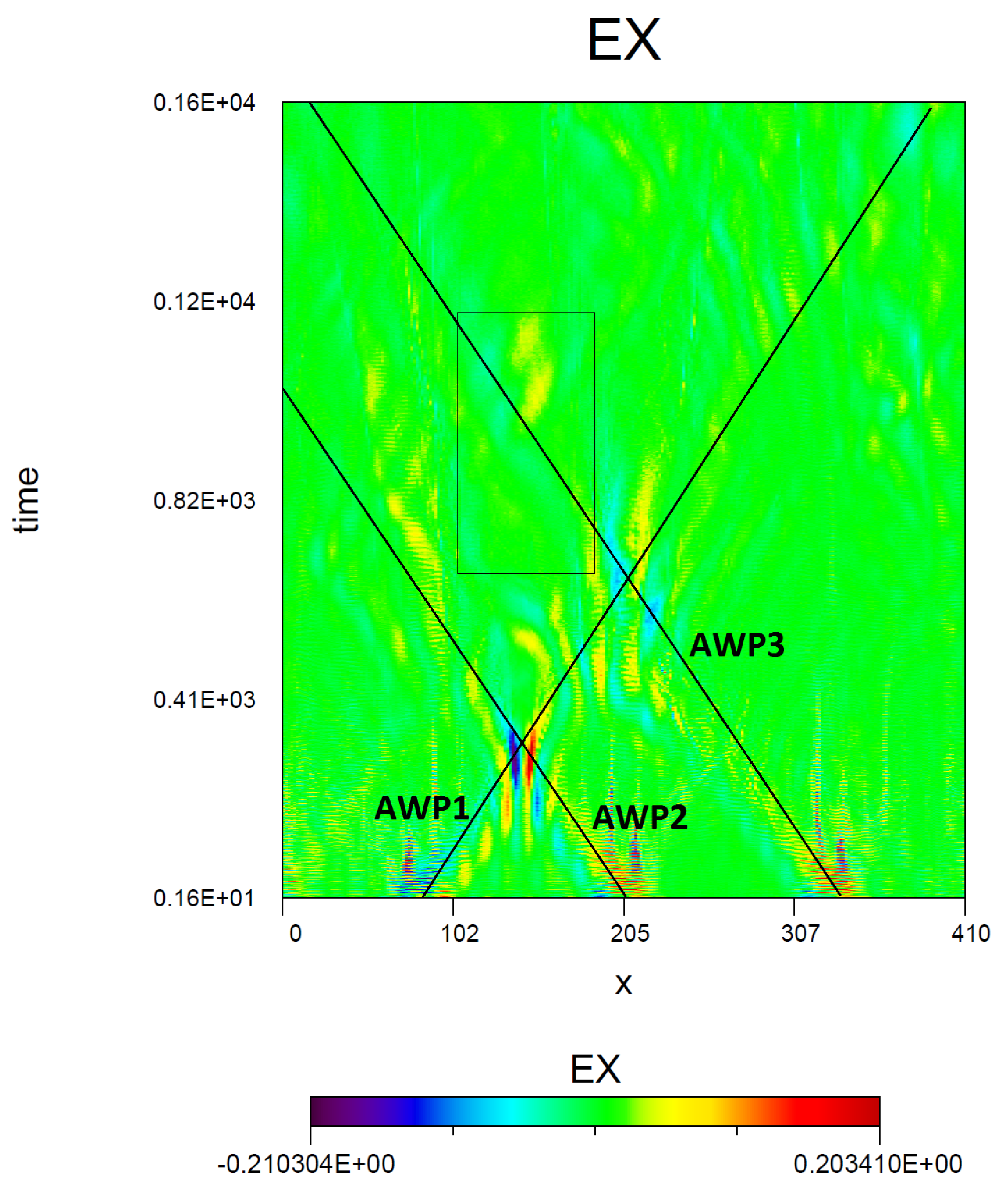}
\hspace{0.03\textwidth}
\includegraphics[width=5.7cm]{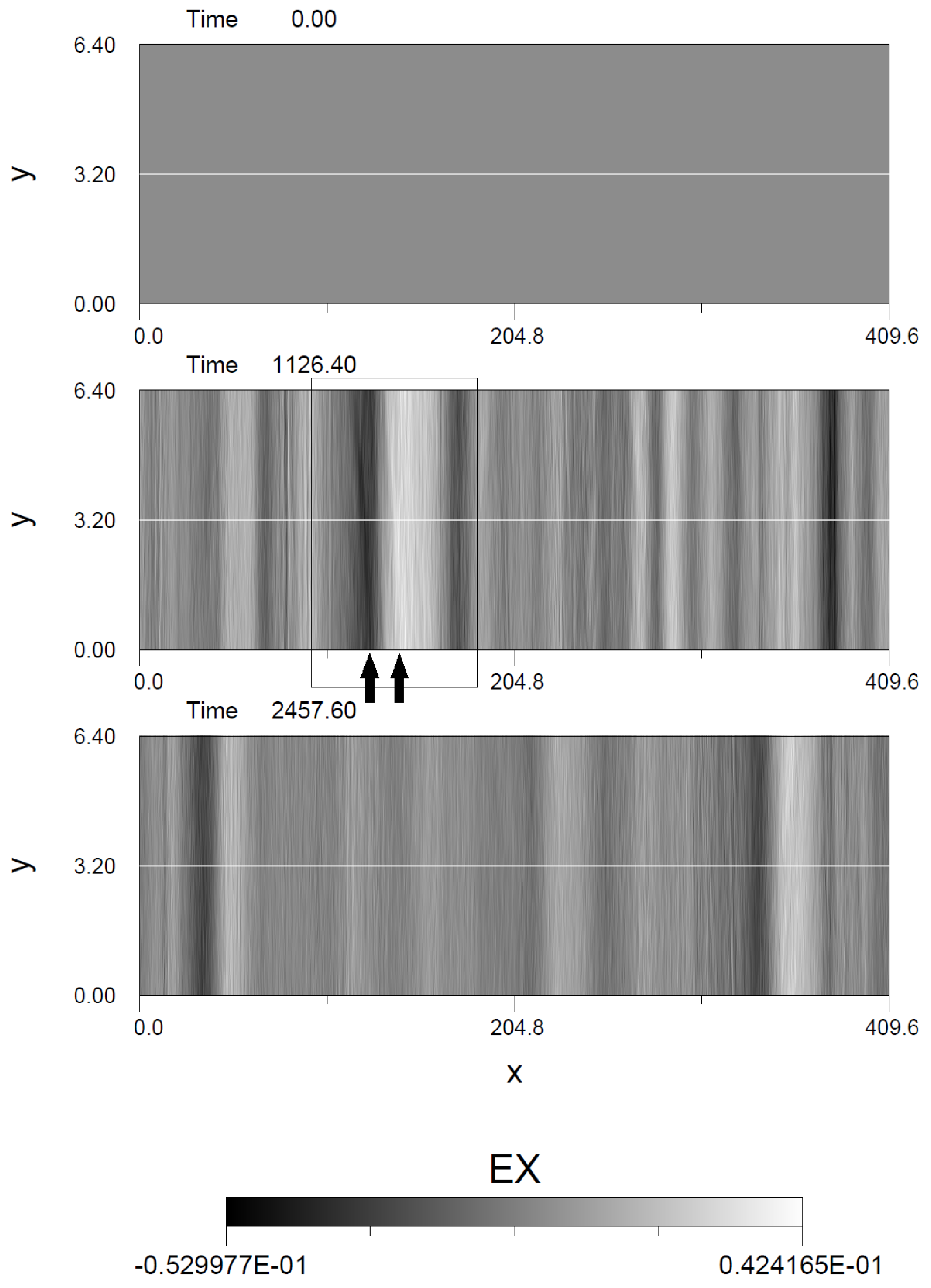}
  \caption{Run (b): The left panel shows the temporal variation of the component $E_X(X,t)$ as a function of the parallel coordinate $X$. The rectangle circumscribes the electric fields that can emerge from AWP3-cavity 1 interaction. The right panel shows snapshots from $t = 0$ to $t = 2457.6$ of the electric field  $EX(X,Y)$. The black arrows show the location of the strong parallel electric fields that arise inside AWP3-cavity 1 crossing region.}
\label{figexb}
\end{figure}

\begin{figure} 
\center
\vspace{0.05\textwidth}
\includegraphics[width=6.4cm]{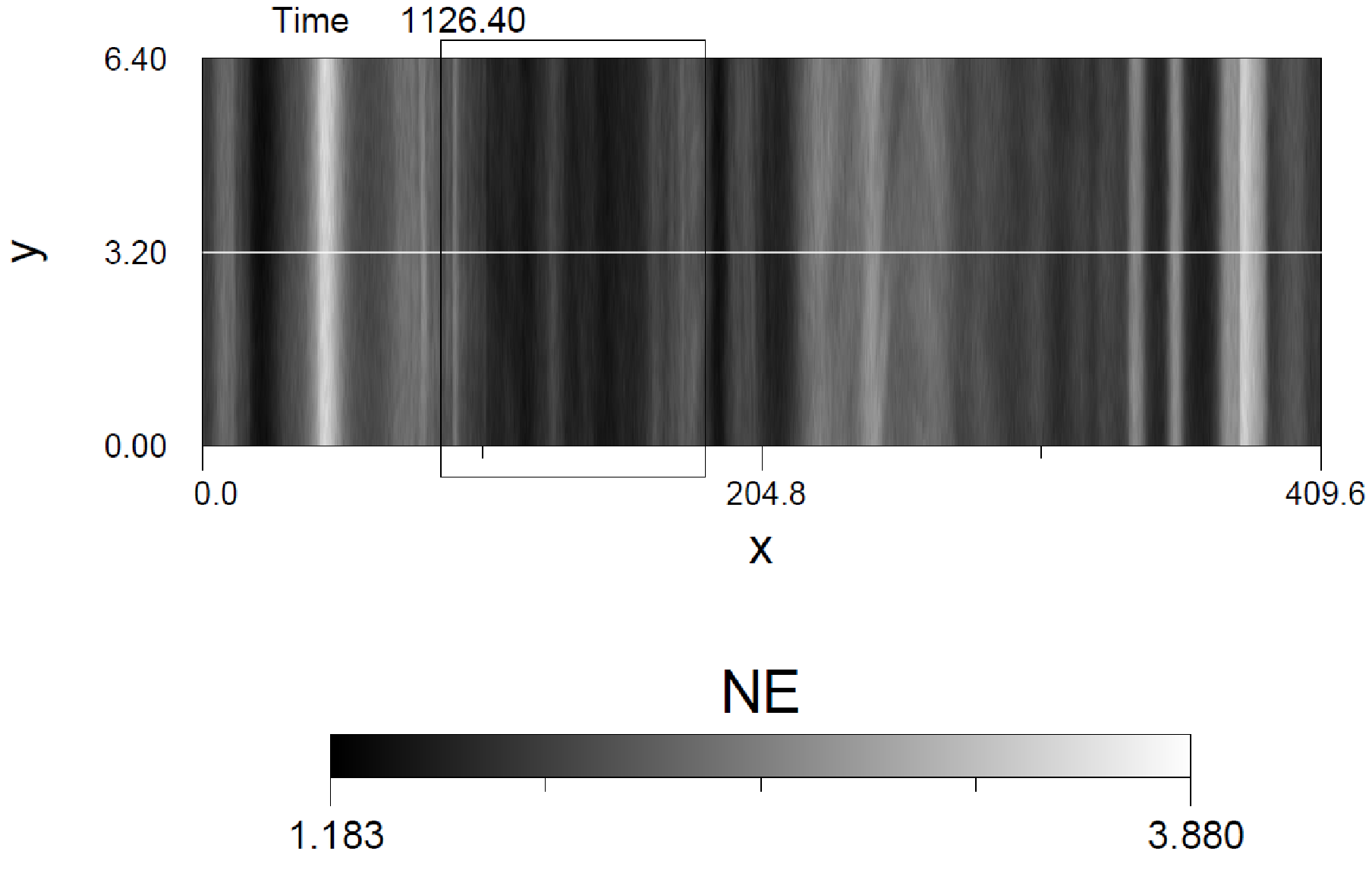}
\hspace{0.03\textwidth}
\includegraphics[width=6.4cm]{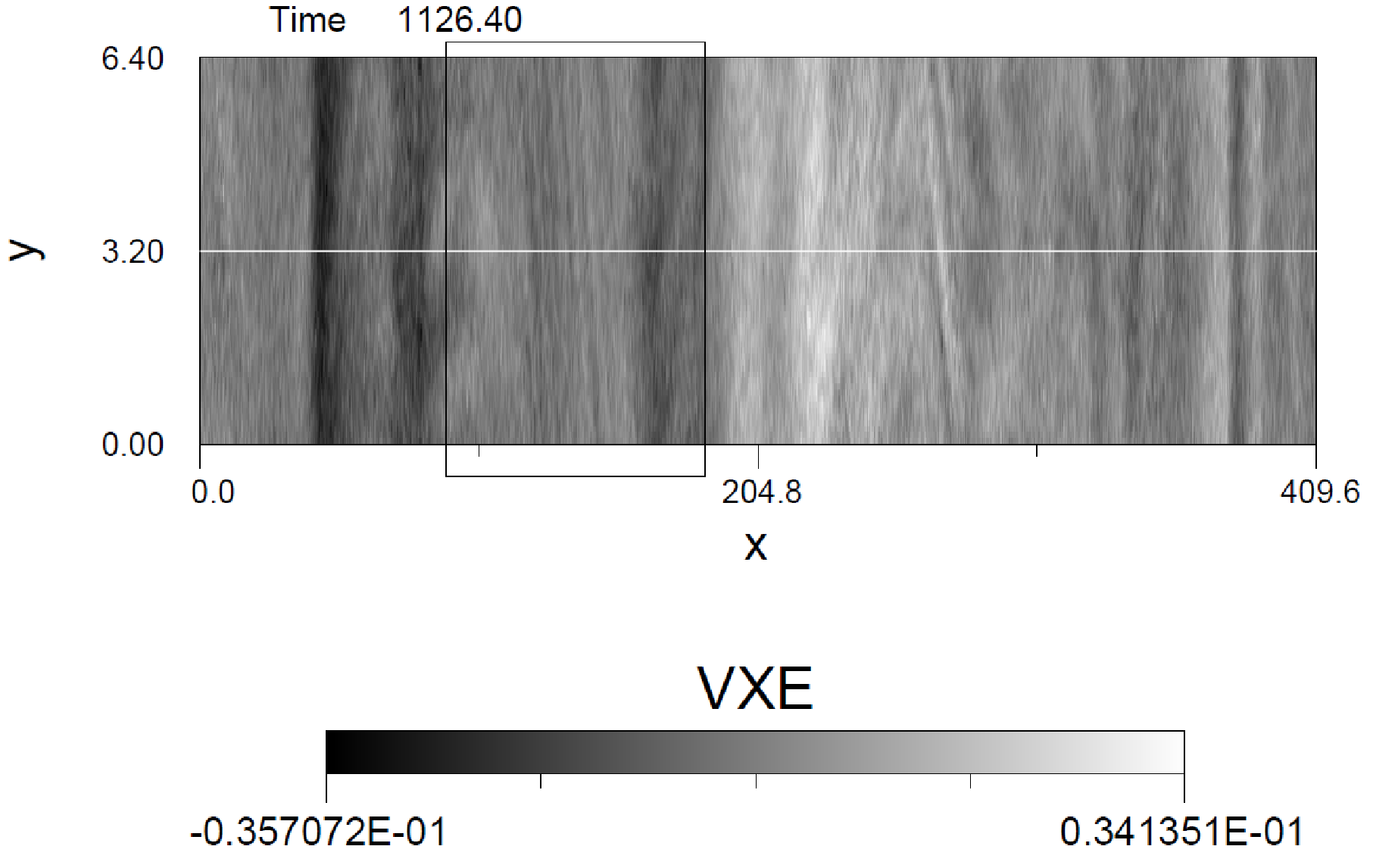} 
  \caption{Run (b): Snapshot $t = 1126.4$ of the electron density $N_e(X,Y)$ in the left panel and the electron parallel velocity $VXE(X,Y)$ in the right panel.}
\label{cne1fig4}
\end{figure}

However, APAWI density gradients $N_e(X,Y)$ show a longitudinal modulation like profile for all our simulations (see  the left panel of Figure \ref{cne1fig4}). The  parallel electric fields observed for the simulations (a) and (b) in Figure \ref{fig2.5} and the right panel of Figure \ref{figexb}, respectively show similar longitudinal profile.

Phase-mixing process in the case of longitudinal gradient of density is more complex. In the context of resonant mode conversion and using analytic calculations, \citet{wu19} showed that kinetic Alfv\'en wave is hardly excited when $\alpha \le 40^{\circ} $, where $\alpha$ is the angle between the background parallel magnetic field and the density gradient of the plasma.  \citet{genot99} demonstrated analytically that a pure parallel electric field is generated in the regions of transverse density gradients, whereas in the regions of longitudinal density gradients, the emerged parallel and perpendicular electric field are coupled.  
 \citet{lysak08} have performed simulations where they considered the case of parallel and perpendicular density gradients in the Earth magnetosphere. They observed that a parallel electric field was developed at the gradient regions in the Alfv\'en speed.

In addition to theoretical possibilities that are cited above, we suggest for our simulation that the passage of the Alfv\'en waves can induce a relative drift $\bf{V_0}=\bf{E} \times \bf{B}/B^2$ along the discontinuity boundary between the background plasma and the longitudinal APAWI density cavity structure. This will generate an interface waves along the discontinuity which allows the incoming wave to interact with a wavy density gradient (oblique gradient). These wavy fiber-structures can be seen in the left panel of Figure \ref{cne1fig4} for the simulation (b). When the wave reaches the oblique density cavity, the distorted wave front develops small-scale $k_{\perp}$ structures in the transverse $Y$-direction. Since we are in inertial regime, when $k_{\perp}^{-1}$ reaches the electron inertial length $c/\omega_{p0}$, a parallel electric field is generated at this transverse gradient by phase-mixing process.

The electrons and ions drift are at the same velocity $\bf{E} \times \bf{B}$. Furthermore, the transverse size of our box $L_Y$ is large in comparison to the ion gyroradius $\rho_i=1.25$. This gives an MHD behavior to the plasma in the transverse direction \citep{fag17}.

In our simulations, interface waves can be generated when the third wave packet cross the region of longitudinal density (cavity 1) that emerged from APAWI process. Two drift terms can contribute to the creation of small-scale  structures ($k_{\perp}^{-1}$) in the transverse direction $Y$: the first one is  $\bf{\delta E_{3Z}} \times \bf{B_0}$, where $\delta E_{3Z}$ is the electric field perturbation of the third AW in the $Z$-direction. The second term is $\bf{E_X} \times \bf{\delta B_{3Z}}$, where $\delta B_{3Z}$ is the magnetic field perturbation of the third AW in the $Z$-direction, and $E_X$ is the parallel electric field that results from the collision between the two initial Alfv\'en waves (APAWI). This last term which depends on the lifetime of $E_X$ vanish rapidly in comparison to the first term.  To compare between the contribution of these two terms we have calculated the ratio of the magnitude of the first term to the magnitude of the second term:

\begin{equation}
\label{ratiodrift}
\frac{\bf{\delta E_{3Z}} \times \bf{B_0}}{\bf{E_X} \times \bf{\delta B_{3Z}}} \sim \frac{V_3 B_0}{E_X},
\end{equation}

where $\delta E_{3Z}=V_3 \delta B_{3Z}$, $V_3$ is the phase velocity of the incoming third Alfv\'en wave, $\delta B_{3Z}/B_0=0.1$ or $0.05$ and $B_0=0.8$. In \citet{mottez15}, the simulation AWC009 is quasi similar to the APAWI part (RH-RH) of the simulation (a) where the maximum of APAWI parallel electric field $E_X$ is about 0.013. In simulation (a), the phase velocitie of the third Alfv\'en wave packet $V_3$ varies from 0.0855 to 0.2029. Then the ratio in equation \ref{ratiodrift} for the simulation (a) is between 5.26 and 12.49. This means that in simulation (a), the first drift term $\bf{\delta E_{3Z}} \times \bf{B_0}$ is almost predominant in comparison to the second one.

It is possible that interface waves are generated along the longitudinal APAWI density structures earlier when the two initial waves (RH-RH) (or RH-LH) move away after their interaction. Thus, outgoing wave packets will cross a part of the emerged APAWI density generating primordial small-scale $k_{\perp}^{-1}$ structures in the transverse direction before the passage of the third wave packet. Given that, it is interesting to calculate the ratio of the velocity drift magnitude ($\bf{\delta E_Z} \times \bf{B_0}$) of a single (RH) Alfv\'en wave to the velocity drift magnitude of a single (LH) wave. This ratio is about 1.16 ($m=1$) and 7.13 ($m=16$) for waves of the same amplitude,  which means that the single (RH) Alfv\'en wave will induce a more important transversal modulations in the APAWI longitudinal density structures than the single (LH) wave of the same amplitude. This result is in favor of stronger parallel electric field generation in simulations (a) and (b). 

In the phase-mixing process, higher incoming wave amplitude can induce more concentration of space charge on the transverse density gradient regions, as well as the creation of a stronger parallel electric fields,  which in turn accelerate electrons more efficiently \citep[see][]{genot99, tsi08}. This is the case of simulation (b) where the amplitude of the third incoming AW is larger compared to that of the  other simulations.

For both simulation (a) and (b), the phase-mixed parallel electric fields associated to longitudinal density gradients shows clearly a wavy fiber-like structures ($ 93 \le X < 205 $ for $t>600$).
In general,  these tiny localized structures are difficult to observe, particularly for snapshots where a poor contrast shows mainly the global longitudinal modulation. They are better discernible in simulation (b) for the electron  parallel velocity $VXE(X,Y)$ (right panel of Figure \ref{cne1fig4}) at $t=1126.4$. Different patterns of these structures can be seen depending on the size of the gradient and the wavelength of the perturbation along the transverse direction $Y$.

In simulation (a), the horizontal length of the small-scale parallel electric associated to phase-mixing process measured from the left panel of Figure \ref{fig2.5} ($t=716.8$) is about  $\delta X \sim 0.98$. In simulation (b), the parallel electric field associated to the acceleration region is rather formed by a cluster of small-scale fiber structures (right panel of Figure \ref{figexb} at $t=1126.4$), where the the horizontal length of a single fiber is about $\delta X \le 0.98$. These sizes fit very well the size of the the electron inertial length  ($c/\omega_{p0}=1$).

In conclusion,  when the AWP cross the longitudinal APAWI density gradient structure, fiber structures start to oscillate in the transverse direction. Parallel electric fields are generated in each small-structures by phase mixing process as long as the size of the fiber-transverse-section crossed by the wave front is in range of  $c/\omega_{p0}$. Then, the large size parallel electric fields associated to the modulations of density is the contribution of all fine-structure parallel electric fields.


\section{Discussion and conclusion}

APAWI process causes a significant modulation of the plasma density and accelerate particles in the parallel direction. This process have been studied for higher and lower ambient magnetic field by \citet{mottez12, mottez15}, respectively.
In the present work we are trying to generalize the APAWI process by interacting a third parallel Alfv\'en wave packet with the two initial ones. The aim was to explain the observation of high-energy electrons through nonlinear wave-wave interaction model.

We have investigated the nonlinear interaction of three crossed parallel Alfv\'en wave packets in initially uniform plasma ($k_{\perp}=0$), by using a particle in cells (PIC) numerical simulation code. The two closest Alfv\'en waves will collide first in the context of APAWI process giving rise to an inhomogeneous plasma at the APAWI crossing region. Then the third Alfv\'en wave interact in turn with the emerged density modulations and waves.
We observe parallel electric fields and ion beams localized at the APAWI crossing region. We suggest that they result from phase-mixing process between the third Alfv\'en waves and the APAWI density depletion. This is the case for waves with different polarizations like (RH-LH LH) or (RH RH-LH) configuration,  where (RH) and (LH) mean the right and  the left circular polarization respectively. However, electron beams ($\sim 5-7 V_{Te}$) associated to larger parallel electric fields are observed for the interaction (RH-RH RH), particularly when large amplitude propagating Alfv\'en waves are simulated.

While APAWI density structures show a longitudinal gradients, we explain the transversal modulations to the propagation of interface waves generated mainly by the velocity drift $\bf{\delta E_Z} \times \bf{B_0}$ along the boundaries of the emerged longitudinal APAWI density structures, where $B_0$ is the background magnetic field, and $\delta E$ is the electric field perturbation of the crossing waves. Our simulations were performed in the inertial regime. Therefore,  the transverse size of the small-scale density gradients $k_{\perp}^{-1}$ have to be of order of $c/\omega_{p0}$ to produce a strong parallel electric field that can accelerate electrons in the direction of the background magnetic field.
We have estimated from simulation (b) that the longitudinal size of the large phase-mixed parallel electric field structure is of order of 6 km in the auroral zone, where the transverse scale length (fine structures)  $c/\omega_{p0}\sim 168$ m. The possibility and the observation of electric field structures of comparable size have been discussed in \citet{karl20}. Nevertheless, large amplitude interface waves can evolve to Kelvin Helmholtz instability. 
In this context, many authors have reported the important role that plays this instability in the Earth magnetosphere dynamics \citep[see][and references therein]{fag17}. Structures and even vortices from this instability have been also observed in the auroral sheet \citep{hallinan70}.

Actually, the process of interaction of three parallel waves invoked in this study is different from the APAWI one. APAWI process involve simultaneous crossing of parallel counter-propagating Alfv\'en waves, whereas in the mechanism studied here, the third wave packet arrives with a very short delay compared to the instant when the two initial waves interact. From a geometrical point of view, the two process are comparable since wave packets propagate in the parallel direction to the background magnetic field.  Nevertheless, to compare between the two process from physical point of view, the three propagating Alfv\'en waves have to cross together simultaneously. This can be possible only if the third wave packet propagates obliquely (or three oblique Alfv\'en waves). However, this involves a further $k_\perp$ from the oblique propagation in addition to $k_\perp$ generated from non-linear coupling of the counter-propagating Alfv\'en waves, which changes a bit the problem. This may be the subject of further study.

It is interesting to notice that because of the periodic horizontal boundary conditions, the complex interaction that happens at the final stage of simulations can give us an idea about phenomena related to multiple reflection of Alfv\'en waves in a resonant cavity. However, how to justify the presence of waves of different polarization in our simulations since basically we want to study the self-interaction of a single Alfv\'en wave packet when it undergoes multiple reflection?
To answer to this question, we can imagine a situation where a (RH) Alfv\'en wave packet propagates in ($\beta < m_e/m_i$) plasma as in our simulation but surrounding by a plasma with ($\beta > 1$). Using a kinetic approach, \citet{buti00} have shown that (RH) Alfv\'en wave packet is unstable for plasma with $\beta > 1$ and can collapse when it arrives at a turning point depending on the initial amplitude of the wave packet. At this critical point, the (RH) wave packet would change polarization to become like a (LH) wave packet. Given that, this transformed (LH) wave packet can return back from a reflective boundary layer or from strong conditions turning point and cross the initial (RH) wave packet. If this interaction (RH-LH) can occur, then three wave packets (or more) of different polarizations (and amplitudes) can interact also under the same circumstances. The mechanism of conversion from fast magneto-acoustic wave to Alfv\'en and slow wave above an active region ($\beta < 1$) in low solar corona, including a change in polarizations and amplitudes,  offers also a possibility to a single wave to collide with their counter-propagating parts after reflections and damping in regions of rapidly Alfv\'en speed increases in open magnetic field \citep{cally12}, or in closed magnetic field configuration like coronal magnetic loop \citep{fletcher08}. Obviously, the wavelengths of the MHD waves and the size of the resonant cavity have to be in the range where kinetic effects of the plasma can take place.

Alfv\'en waves propagating through a longitudinal density gradient can undergo a reflection since it propagation velocity $V_A(X) =B_0/\sqrt{\mu_0 \rho_m(X)}$ changes along the parallel direction. \citet{musi92} have calculated the critical frequency $f_c$  under which incident Alfv\'en wave can be reflected by a longitudinal gradient. In term of our dimensionless variables, this frequency can be written as $\Omega_{c} = \frac{1}{2} V_{A0} \sqrt{(V_A')^2 +|2V_AV_A''|}$, where the prime in this equation indicates the spatial ($X$) derivative, and $V_{A0}=0.08$ is the Alfv\'en velocity at the background uniform plasma $N_{e0}$.
The Alfv\'en velocity $V_A(X)$ for a single wave is calculated along a typical APAWI cavity which we approximate with a longitudinal Gaussian density profile. We assume that the maximum amplitude of a typical moderate cavity and it width are: $Max(N_e(X)/N_{e0})=0.4$, $\delta X=18$, respectively.  We have find that $\Omega_c \approx 1.4 \times 10^{-3}$. From initial conditions, the frequencies in the (LH) wave packet varies from $1.13 \times 10^{-3}$ ($m=1$) to $7 \times 10^{-3}$ ($m=16$), and the frequencies in the (RH) wave packet varies from $1.31 \times 10^{-3}$ ($m=1$) to $5 \times 10^{-2}$ ($m=16$), where $m$ is the wave number. Therefore, it is clear that a small part ($m=1$) of the wave packet will undergo a reflection from APAWI longitudinal density cavities. However, the equation for $f_c$ in \citet{musi92} is valid only for linear Alfv\'en waves of relatively low amplitude and for non-fibril cavity. Furthermore, Alfv\'en wave packet modulates in time the density gradient making it profile time-dependent, which makes the calculation of the dynamical $f_c$ more complicated. 

In addition to the non-linear phase mixing process, it is possible that the reflection of the wave packets from the emerged longitudinal APAWI density gradients contributes to electron heating, it depends on the transmitted wave amplitude \citep{bose19}.  It is possible also that a part of the parallel Alfv\'en wave can be trapped between two reflective longitudinal APAWI gradients if it wavelength is smaller than the distance between these two gradients. At the MHD scale, \citet{moore91} have shown that the heating of coronal holes is predominantly caused by the reflection and the dissipation of the trapped Alfv\'en waves rather than of the transmitted waves. By invoking turbulence cascade, similar mechanism could possibly occurs at the kinetic level. However, it is difficult to confirm these possibilities from our simulations since we observe the contribution of three propagating Alfv\'en waves. 
In this regard it will be important to carry out simulations of a single parallel Alfv\'en wave propagation through longitudinal (fibril) density gradients of different sizes. These will be the topics of a forthcoming investigation.

\section*{Acknowledgements}
I would like to express my sincere gratitude to Fabrice Mottez from Observatoire de Paris for the original idea of the paper, and for years of discussion, inspiration and help. I am also thankful to the anonymous referee for constructive comments and suggestions that improved greatly the quality of the paper.
\section*{Declaration of Interests}
The authors report no conflict of interest.
\section*{Data availability}
The data that support the findings of this study are available within the article.



\end{document}